\documentclass[journal]{IEEEtran}
\ifCLASSINFOpdf
\else
\fi

\usepackage{booktabs} % table (toprule, ...)
\usepackage{color} % colorful text
\usepackage{multirow} % multirow in a table
\usepackage{graphicx}
\usepackage{amssymb,amsmath,bm,graphicx,url}
\usepackage{textcomp}
\usepackage{subcaption}
\usepackage{enumitem}
\usepackage[noadjust]{cite}
\usepackage[activate={true,nocompatibility},final,tracking=true,kerning=true,spacing=true,factor=1100,stretch=10]{microtype}
\usepackage{pgfplots}
\usepackage{array}
\usepackage{caption,subcaption}
\usepackage{xcolor}
\newcolumntype{P}[1]{>{\centering\arraybackslash}p{#1}}
\newcolumntype{M}[1]{>{\centering\arraybackslash}m{#1}}

\begin{document}

\title{Realizing Petabyte Scale Acoustic Modeling}

\author{Sree Hari Krishnan Parthasarathi, Nitin Sivakrishnan, Pranav Ladkat, Nikko Strom% <-this % stops a space

\thanks{All authors are employed by Amazon.com, Inc., \{sparta\textbar{}nitins\textbar{}ladkat\textbar{}nikko\}@amazon.com}% <-this % stops a space
}

%                     ^------------------^------------------^------Do not want these spaces!

\markboth{Journal on Emerging and Selected Topics in Circuits and Systems}%
{A \MakeLowercase{\textit{et al.}}: Realizing Petabyte Scale Acoustic Modeling}

%\IEEEpubid{0000---0000/00\$00.00~\copyright~2015 IEEE}


\maketitle

\IEEEpubid{\begin{minipage}{\textwidth}\ \\[12pt] \centering
Digital Object Identifier 10.1109/JETCAS.2019.2912353 \\
  2156-3357 \copyright 2019 IEEE. Personal use is permitted, but republication/redistribution requires IEEE permission.\\
  See http://www.ieee.org/publications standards/publications/rights/index.html for more information.
\end{minipage}}

\IEEEpubidadjcol

\begin{abstract}
Large scale machine learning (ML) systems such as the Alexa automatic speech recognition (ASR) system continue to improve with increasing amounts of manually transcribed training data. Instead of scaling manual transcription to impractical levels, we utilize semi-supervised learning (SSL) to learn acoustic models (AM) from the vast firehose of untranscribed audio data. Learning an AM from 1 Million hours of audio presents unique ML and system design challenges. We present the design and evaluation of a highly scalable and resource efficient SSL system for AM. Employing the student/teacher learning paradigm, we focus on the student learning subsystem: a scalable and robust data pipeline that generates features and targets from raw audio, and an efficient model pipeline, including the distributed trainer, that builds a student model. Our evaluations show that, even without extensive hyper-parameter tuning, we obtain relative accuracy improvements in the 10 to 20$\%$ range, with higher gains in noisier conditions. The end-to-end processing time of this SSL system was 12 days, and several components in this system can trivially scale linearly with more compute resources.

\end{abstract}


% \fi
\IEEEpeerreviewmaketitle

\section{Introduction} \label{Introduction}
\IEEEPARstart{M}{odern} machine learning (ML) relies on large-scale data; for hard problems more data consistently lead to better models. In the field of automatic speech recognition, there is a well known maxim: {\it there is no data like more data}~\cite{jelinek:2004}. This has naturally led to the use of ever increasing amounts of speech data. From tens of hours of speech (TIDIGITS~\cite{tidigits:1993}, TIMIT~\cite{timit:1993}, WSJ~\cite{wsj:1992}), to hundreds (Switchboard~\cite{switchboard:1992}), and thousands (Fisher corpus~\cite{fisher:2004}). Recently, training data sizes on the order of ten thousand hours of speech are not unusual (\cite{deepspeech2:2016}, \cite{gong-semi:2016}), and while building an AM from a hundred thousand hours has still been uncommon, ~\cite{soltau:2016} showed that increasing from several thousand hours to a hundred thousand hours of lightly supervised speech data can improve speech recognition accuracy significantly.

This increase in the amount of data in ASR and other ML fields requires ever more efficient data and ML processing pipelines. A rich infrastructure has emerged -- commercially and in Open Source communities -- to serve both hardware and software requirements of large-scale ML. There has been extensive developments in powerful ML toolkits (Spark MLLib, mlpack, Scikit-Learn), ML cloud services (AzureML, Amazon ML, Google Cloud Machine Learning), distributed storage (S3, HDFS ~\cite{hadoop2010}, GoogleFileSystem), frameworks for distributed compute (Hadoop ~\cite{hadoop2010}, Spark ~\cite{apachespark2016}) and Deep Learning (PyTorch ~\cite{pytorch2017}, MxNet ~\cite{mxnet2015}, TensorFlow ~\cite{tensorflow2015}). These tools make it easy to train ML models. However, for the largest, most data intensive ML systems it is critical to use generic tools efficiently, and sometimes develop custom solutions, to avoid memory, bandwidth or processing bottlenecks. In this paper we present our end-to-end ML system for building an acoustic model based on 1 Million hours of speech in 12 days. This is one order of magnitude more speech data than has been reported in any previously published work in ASR ~\cite{soltau:2016}.

While many of the techniques and tradeoffs are applicable to other fields, we demonstrate the large-scale challenge here by tackling acoustic model (AM) training for ASR. The AM in the ASR system takes a sequence of acoustic feature vectors as input and produces a sequence of phonetic probability densities as output~\cite{hintondeep:2012}. At each time step, it produces a posterior probability for each phonetic class. Because of the large variability in speaker characteristics, dialects, accents, and acoustic environments, producing a highly accurate AM requires large amounts of speech data, and the model is typically trained as a phonetic classifier with supervision from annotated speech data. Thus, for each training speech utterance, a human annotator listens and assigns the correct spoken words. However, at the scale of 1 Million hours of speech, human annotation becomes impractical. Both the cost and the logistics required to manage such a large undertaking in a reasonable amount of time are prohibiting. Therefore, here we use semi-supervised training, where only a fraction of the speech data is annotated.

To characterize the scale of the task, 1 million hours is equivalent to a constant stream of speech, 24/7, for 114 years -- far more than any human will hear in a lifetime. The number of utterances is on the order of one billion, and since we extract hundreds of feature values for every 30 milli second of each utterance, our models process over 1 trillion attributes during training. In terms of raw size for storage, for the commonly used 256 kbps wav formats for audio, 1 million hours of audio uncompressed requires 107 TB of disk space for storage. Building a robust data transformation and training pipeline that handles such data sizes has required close attention to, among other considerations, the failure modes of underlying software, machines, disks and networking channels; limitations of cluster computing frameworks; theoretical algorithmic scaling bottlenecks and monetary cost of resources utilized.

\IEEEpubidadjcol

For the SSL system discussed in the paper, we employed the student/teacher learning paradigm, specifically focusing on the student learning subsystem. We factored the student learning subsystem into two pipelines: data preparation and model training. Each has unique scaling challenges. Aside from managing a cluster of compute nodes and distributed data storage, the data preparation step performs shuffling and normalization that is non-trivial in a large scale distributed system. The model training on the other hand must address the challenge of large-sale data-parallel neural network training. This is a fast evolving field that has received significant attention in recent years (\cite{strom2015scalable}, \cite{1bit-sgd}, \cite{qsgd}, \cite{deepcompression}, \cite{adacomp2017}, \cite{bmuf-2016}).

The remainder of the paper is organized as follows: In the next section we give an overview of the system, followed by section \ref{ML Training Infrastructure} about the computing infrastructure. In sections: \ref{SSL Data pipeline}, \ref{SSL Model pipeline}, and \ref{Distributed Training} we give more details about the most prominent components of the system, and finally section \ref{Results} reports our results and \ref{Conclusion} concludes.

\section{Overall System} \label{Overall System}
\subsection{Background on the ASR System}

\subsubsection{Signal processing system}
The Alexa family of devices~\cite{RohitTalk2015} use an array of microphones arranged in different geometries. The signal processing system takes the raw channels and the playback channel as input and returns a single signal for downstream processing by the wake word detector (\cite{panchapagesan2016multi}~\cite{maas2016anchored}) and the ASR system~(\cite{parthasarathi2015fmllr},~\cite{garimella2015robust}). This system uses beamforming (BF) algorithms to emphasize speech from a desired direction while suppressing audio interference from other directions, and an acoustic echo canceler (AEC) to remove the playback channel from each of the beams.

\subsubsection{Acoustic and Language Models}
The input to the ASR system used in this work is an audio signal, from which we derive a sequence of fixed size acoustic vectors ($X_{1:T} = x_1,...,x_T$). The ASR system solves the problem of finding the most likely sequence of words ($W_{1:M_W} = w_1,...,w_{M_W}$) given the sequence of acoustic vectors
\begin{align}
\label{am:asr}
\hat{W} &\propto {\arg\max}_{W} \underbrace{p(X|W; \Theta_{\text{AM}})^{{\kappa}}}_\text{AM} \underbrace{P(W;\Theta_{\text{LM}})}_\text{LM},
\end{align}
where $\Theta_{\text{AM}}$ and $\Theta_{\text{LM}}$ are the free parameters of the acoustic and language model, and $\kappa$ balances the impact of the acoustic model against the language model. The AM is based on the standard \textit{HMM/deep learning} hybrid, and we summarize details relevant to this paper in Section~\ref{sec:sys_description}. Other aspects of this system have been described elsewhere~(\cite{parthasarathi2015fmllr},~\cite{garimella2015robust},~\cite{king2017robust},~\cite{minhua2019icassp}).

The LM~\cite{ankur2018lm} estimates the \textit{a priori} probability that the speaker will utter a sequence of words. The decoder is an FST based decoder using an optimized dynamic composition approach. In this work, we use a large n-gram model.

\subsection{Fully Supervised Acoustic Model}
\label{sec:sys_description}
We use an HMM-LSTM hybrid. The HMM models low-frame rate single state triphone units~\cite{pundak2016lower}. States are clustered down to 3,183 senones using phonetic decision trees. The acoustic features consist of 64-dimensional log mel-warped energies computed on audio signals every 10 ms with a 25 ms analysis window. These are stacked three at a time and sub-sampled to a 30 ms advance. A causal mean estimate is computed and subtracted, and finally global mean and variance normalization is applied. To compensate for sub-sampling, features are created at three different offsets for each utterance.

The LSTM model is a stack of five layers, each consisting of 768 units resulting in about 24 M parameters. The model has a three-frame look-ahead. The training data is 7,000 hours of labeled US English data. The models are trained first with the cross-entropy criterion (CE), using alignments computed on the labeled data. First, we follow an exponential learning rate decay for ten epochs, with chunked BPTT for greater parallelization efficiency~\cite{doetsch2014fast}. In this technique, utterances are split into smaller sub-sequence chunks (here, 32 frames) and the sub-sequences are randomized. For each epoch we cycle through a different feature offset. Then the models are fine-tuned using full sequence CE BPTT for two more epochs. Finally, three epochs of the sequence discriminative criterion state-level minimum Bayes risk (sMBR) is applied ~\cite{kingsbury2009lattice}.

\subsection{Semi-Supervised Learning}
SSL has a long history in ASR (\cite{kemp1999unsupervised},~\cite{lamel2002lightly},~\cite{ma2006unsupervised}). Self-training is the most commonly used approach where typically there is a smaller labeled dataset, and a much larger unlabeled dataset. The labeled data is used to train a seed model from a powerful model family, which is used to decode the unlabeled data at the second stage (often large beam sizes are used). The most reliable hypotheses are selected based on confidence measures~\cite{siu1997improved} and the speech data with the selected hypotheses are used for re-training the AM.

\begin{figure}[htb]
 \centering
\includegraphics[width=3.5in]{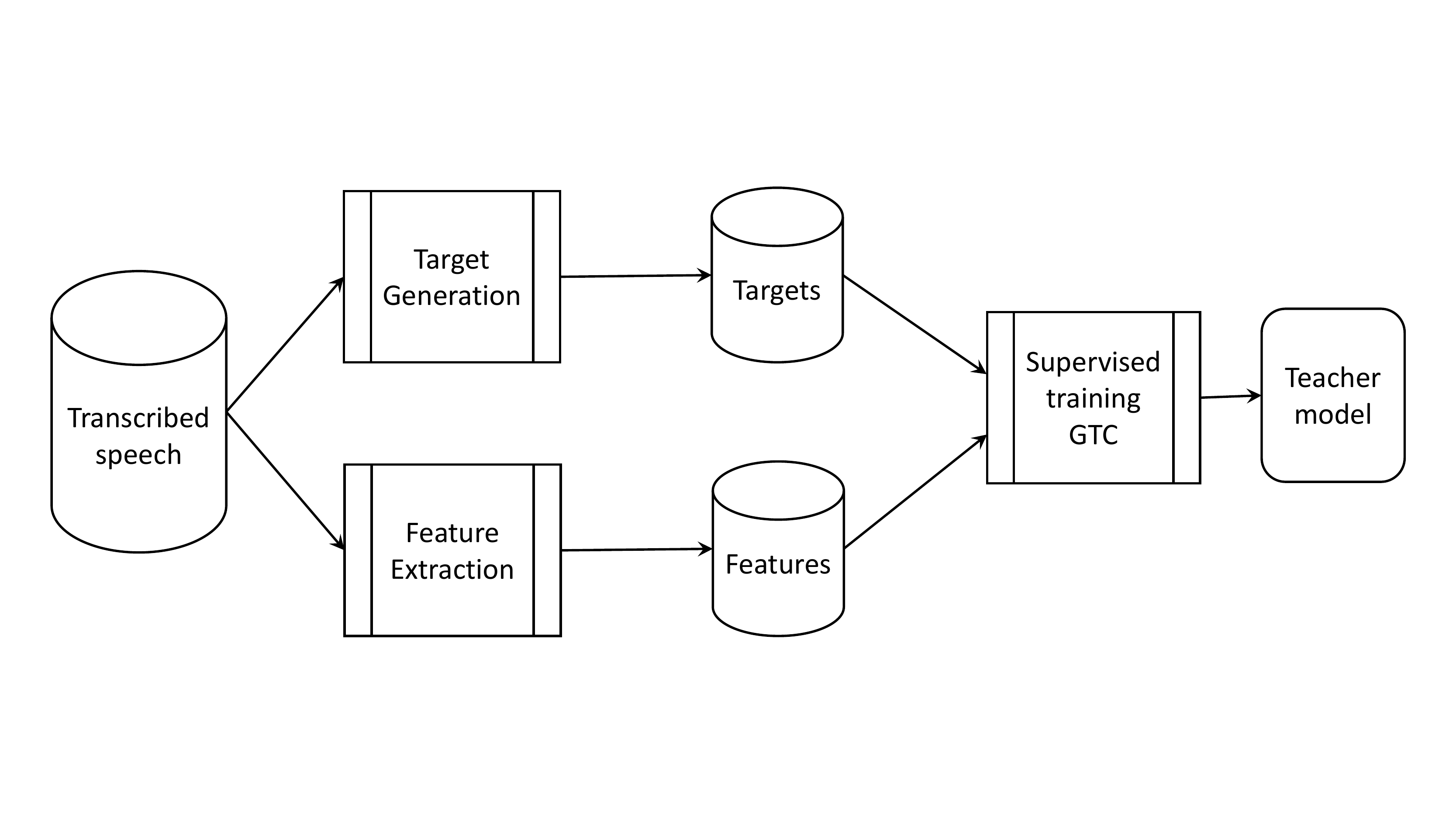}
 \caption{\textit{SSL Teacher Subsystem: Training the teacher model on transcribed data. We also show the distributed trainer ``GTC'', of which we will discuss more in Section~\ref{Distributed Training}.}}
\label{fig:teacher_system}
\end{figure}

\begin{figure}[htb]
 \centering
\includegraphics[width=3.5in]{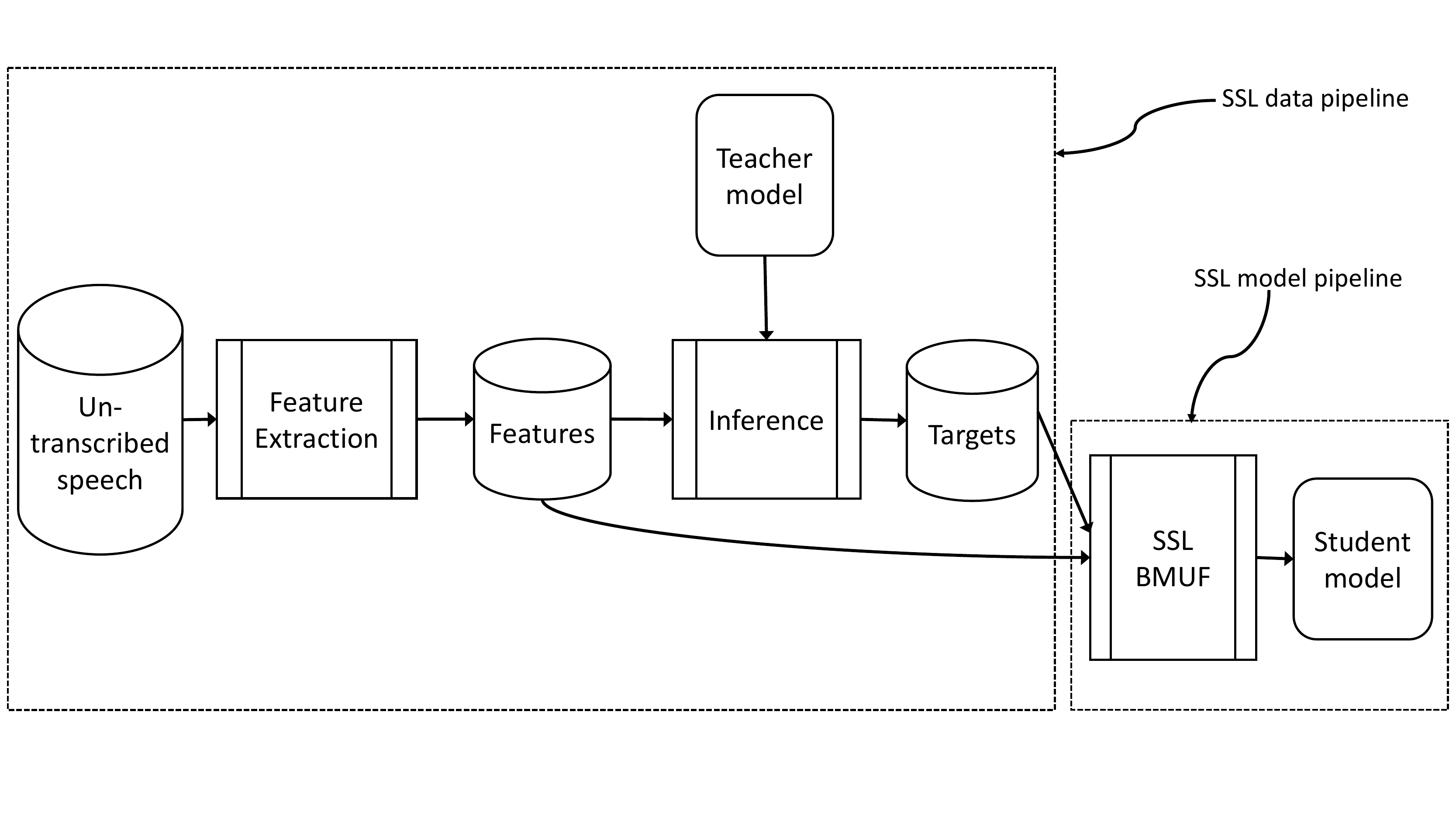}
 \caption{\textit{SSL Student Subsystem: Training the student model on 1 Mhr of untranscribed data using student/teacher methodology. This figure also illustrates how the data and model pipeline subsystems fit in the overall system. Also shown in the figure is the distributed trainer ``BMUF'', of which more will be described in Section~\ref{Distributed Training}. }}
\label{fig:ssl_system}
\end{figure}

\subsubsection{Overview of the SSL System}
\label{sec:ssl_ts_overview}
Our approach to SSL was to employ the student/teacher learning paradigm, which avoids explicitly modeling confidence scores, thus taking the ASR decoder out of the SSL recipe. The full teacher and student subsystems are shown in Figures~\ref{fig:teacher_system} and ~\ref{fig:ssl_system}. The teacher and student models output probability distributions over senones; the learning objective optimizes the CE loss between these two distributions. The teacher model is not bound by the same constraints as the runtime system where the student will be deployed. Therefore, we can use a more powerful model for the teacher. In our case we used a bidirectional LSTM model which has access to both the past and future audio. Note that this teacher model is more accurate than a production system where live audio is streamed and we cannot use information from the future. We can also use a larger teacher model without incurring compute cost or latency in the live production system. Apart from the difference in model family, the training of the teacher on the labeled data follows the same recipe as the regular LSTMs, discussed in Section~\ref{sec:sys_description}.

We experimented with many different ways to utilize both the annotated and un-annotated data. The final training recipe is periodically inserting the annotated data to the student training interspersed with the un-annotated data. We refer to this method as \textit{scheduled learning}. However, we use only the annotated data in the last steps of the training recipe. Those steps use sequence training which is more sensitive to label errors. A fuller discussion of these modeling challenges and the solutions is presented in~\cite{million2019icassp}, and we summarize this in Section~\ref{sec:scheduled_learning}.

The focus of this paper is a description of the system level challenges in realizing this large scale system. Over the next few sections, we summarize these challenges, outline the resources at our disposal and elaborate the design tradeoffs.

\subsubsection{SSL System Challenges}
In Section~\ref{sec:ssl_ts_overview} we discussed the factorization of the SSL system into teacher and student pipelines. The actual realization of the system is more involved; for simplicity of discussion we factor it into two main components: data pipeline and model pipeline. Each has their own scaling and efficiency challenges.

\textit{SSL Data Processing Challenges}
The major data processing challenges for the SSL pipeline are scalability and robustness. Some computing steps in AM data preparation and training are inherently sequential, and others require cluster inter-worker data transfer; these introduce delays in the processing system. Another challenge is the granularity of a computing step: smaller steps increase the degree of parallelism, but fails to take advantage of joint optimization between steps. In addition for large scale systems, with source data over 100 TB (alternatively, a billion audio files), and with tasks distributed over several thousand computing cores, the system must be resilient against temporary host and process failures, data corruption and network timeout issues. Finally, to make the system viable in practice, it is necessary to do so within tight time and infrastructure budgets.

\textit{SSL Modeling Challenges}
For the modeling subsystem, accuracy and scalability are the most important challenges. From an accuracy standpoint, data selection and filtering for SSL is a question to consider; several sampling strategies have been previously suggested~\cite{pylkkonen2016optimizing}. In addition, SSL with self-training requires good confidence measures, with several previous proposals~(\cite{siu1997improved}, \cite{huang2013semi}). Another challenge with models which have high memorization capability such as LSTM AMs is that label quality becomes even more important~\cite{gong-semi:2016}. A further challenge is applying sequence discriminative training, where label errors have a larger detrimental effect (\cite{manohar2015semi}, \cite{huang2008maximum}). From a scalability standpoint, for the scale of data we consider in this paper, an efficient inference mechanism to generate training targets is an important constraint. Also the model training must address the challenge of large-sale data-parallel neural network training.

\section{ML Training Infrastructure} \label{ML Training Infrastructure}
In this section we describe our training infrastructure in terms of compute, storage, and the ML software.

\subsection{Elastic Computing}
Amazon Elastic Compute Cloud (Amazon EC2) is a web service that provides secure, resizable compute capacity in the cloud. Our pipelines were deployed on EC2 instances. EC2 provides a wide selection of instance types optimized to serve different types of workloads. Instance types comprise varying combinations of CPU, memory, storage, and networking capacity and give the flexibility to choose the appropriate mix of resources for our jobs. We used three different instance types across our pipelines depending on the characteristics of the task:

\begin{itemize}
\item P3: p3.16xlarge hosts are General purpose GPU instances. The instance type hosts 8 NVIDIA Tesla V100 GPU devices per worker. We used P3s for steps that could take advantage of them, such as model training and inference to generate targets.
\item X1: x1.32xlarge machines are dense compute instance types, with 128 cores in their Intel Xeon E7-8880 v3 processors and up to 1,952 GB of DRAM-based memory. We used these hosts for computing steps that have significant data transfer between workers in a cluster. Using X1s reduces the amount of data being exchanged across the network between the workers.
\item C4: c4.4xlarge instance types have 16 Intel Xeon Platinum processors and were used for highly parallel and independent steps in the pipeline, where computing speed is the main concern.
\end{itemize}

\subsection{Distributed Storage}
AWS S3 is a highly available, durable, and scalable key-value data store that does not pose practical limitations on object sizes. S3 was used as the distributed store for audio, features and targets produced by the pipeline, and for the teacher and student models during training. We use three APIs/operations to access objects in S3 throughout our pipelines: \textit{PUT Object}, \textit{GET Object}, and \textit{HEAD Object}. Our data layout and cluster configuration were driven by the following considerations:
\begin{itemize}
\item S3 has an overhead of around 200 milliseconds on reused connections for the three operations.
\item Provided that we are under the network throughput limit of an instance type, and S3 buckets are appropriately partitioned, the average bandwidths for \textit{PUT Object} and \textit{GET Object} operations are 50 MB/s.
\end{itemize}

\subsection{Cluster Computing}
Apache Spark~\cite{apachespark2016} is an open source cluster computing framework that provides a concise programming model for processing large datasets with implicit parallelism and fault tolerance. Central to the Spark programming model is the concept of a Resilient Distributed Dataset (RDD) -- a distributed collection of records spread over many partitions. Spark provides abstractions to operate on RDDs and the unit of parallelism is an RDD partition. A common pattern when using Spark is to integrate data loading and staging from distributed file system implementations. However, we used Spark for its fault tolerant model in our iterative algorithms, but relied directly on S3 to persist data for jobs. In our Spark jobs, an RDD starts with a list of objects in S3 that serve as the unit of parallelization. An alternative would have been to use an S3 based implementation of Hadoop's distributed file system interface, which would internally call different S3 operations to load records from S3 files into the RDD. However, to scale effectively, relying on Spark and S3 directly was a early design decision in this work. There are a few fundamental transformations permitted on RDDs such as map, filter, join, and reduce. A key observation here is that operations that perform a shuffle of the elements or involve a repartition of the RDD, involve data movement across workers and cause bottlenecks when working with large datasets.

For scheduling Spark jobs we use the stand-alone Spark default FIFO scheduler: by design our files were of roughly equal size and we have a far greater number of files than CPU cores; so, the default scheduler works well. However, the default scheduler introduced complexity in target generation where we need to schedule tasks on GPUs, instead of  on CPUs. For such tasks we implemented a custom scheduler.

\subsection{Machine Learning Software}
We made use of the open source ASR toolkit, Kaldi~\cite{Kaldi2011} for extracting features from audio, computing and applying normalization techniques, and to serialize data derived from utterances. We use an in-house distributed deep-learning training toolkit~\cite{strom2015scalable} for training the acoustic models that has been optimized for the task. We investigated two types of distributed training: Gradient Threshold Compression (GTC)~\cite{strom2015scalable} and Blockwise Model Update Filtering (BMUF)~\cite{bmuf-2016}. We discuss this in more detail in Section~\ref{Distributed Training}.

\section{The Data pipeline} \label{SSL Data pipeline}
For building acoustic models in the large SSL framework, features computed from the audio and their corresponding machine generated targets need to be generated in a scalable and robust fashion. This involves composing a sequence of several highly scalable steps, which we refer to as the data pipeline. Spark provides resiliency and fault tolerance needed to run some of these steps whose execution time span days. Further, we use S3 as the backing store for some of the intermediate artifacts. We built the pipeline iteratively to aid rapid profiling, fine-tuning, and experimental turnaround. The pipeline is designed to be elastic so that it can scale to even larger datasets.

\subsection{Design Principles}
For a system that operated at this scale, the design was a crucial element. The design of the pipeline follows these principles.
\subsubsection{Optimize input files for parallelism early in the pipeline}
Our first step determines how many files the pipeline will consume, and how data will be organized within the files to optimize processing in the future steps. For example, by aggregating and sharding audio based on speaker early in the pipeline, we were able to avoid costly data transfer in later steps, such as in the feature normalization by speaker algorithm that occurs during feature extraction. The trade-off involved in fixing the file partitions early is that we limit the flexibility of reusing intermediate data artifacts for other experiments and pipelines.
\subsubsection{Avoid distributed algorithms that have high inter-node I/O}
Spark group-by, join, and shuffle operations are I/O intensive, requiring heavy data transfer between nodes in the cluster and we avoid them where possible. Consequently, we do not perform utterance level shuffle operations to provide feature randomization as we had run into its scaling limitations even when working with smaller data sets. Instead we shuffle elements hierarchically during feature file generation. This results in features being less uniformly randomized across our all our files. However we did not observe a degradatation in accuracy improvements when increasing the sizes of our datasets in our experiments using the heirarchical shuffle.
\subsubsection{Aggregate to nearly equal sized files}
Working with larger aggregated files reduces the total number of files; this reduces the number of S3 interactions each of which has an unavoidable latency overhead. The \textit{nearly equal file size} property ensures that Spark's default FIFO job scheduler provides a near optimal processing time. We note that files being our unit of parallelization, reducing the total number of files results in the us limiting the ceiling of parallelization, but in practice we are not close to this limit.
\subsubsection{Perform mini-batch operations locally in a CPU/GPU}
Most pipeline steps rely on invoking local, external processes per CPU/GPU on audio data to perform a transformation function. By operating on mini-batches instead of individual data elements, we reduce resource contention and amortize the process start time over the mini-batch. In addition, mini-batches on GPUs take advantage of low level GPU parallel primitives giving a further performance boost. The trade-off with batches is the increased software complexity in implementing batch interfaces and handling failures during processing of individual elements in a batch.
\subsubsection{Consolidate steps}
At this scale, the aggregate S3 operations for a pipeline step can take substantial time. By reducing the total number of steps, we eliminate entire file set transfers, and in turn reduce end-to-end time. However, this involves a cost trade-off in merging steps. If steps that can leverage cheaper compute are merged with a step that requires more expensive compute, the entire set of merged steps may now need to execute on the more expensive compute nodes. Still, we collapsed several steps in feature extraction to a single step with significant savings.

\subsection{Steps in the Data Pipeline}
\label{sec:dp_steps}
Following the above principles, we implemented a pipeline to produce the data set for the 1 M hour training. It consists of the following steps.
\subsubsection{Data selection}
Prior to any job being run, all potential audio files are stored as individual files in S3. Metadata associated with these individual utterances such as an anonymized speaker id, duration, creation timestamp, locale, and the S3 location of the utterance files are stored as JSON records in aggregated files in S3. The aggregated files are partitioned vertically on time interval and horizontally on speaker id. Since our storage systems could have utterances whose duration in aggregate would be more than a million hours, our first step was to select utterances for training. We followed a simple strategy for this across all data: over the period spanning several months, we uniformly selected 1 million hours of audio. The random selection is performed by first loading the entire utterance metadata set for the duration into an RDD followed by sampling records in the RDD using a mild oversampling strategy based on average utterance length statistics. Once selected, we stored relevant metadata in S3. The output files from this step were stored as nearly equal 200,000 files in S3 (about five hours each). Data from a speaker is grouped into the same file; we retain this partitioning scheme for the rest of the steps in the pipeline. Although this is a step that does need inter-node data transfer for the grouping, the total data being handled is much smaller (200 GB uncompressed); it fits into a single X1 instance and poses no immediate scaling bottleneck.
\subsubsection{Feature extraction}
\label{sec:dp_featureextraction}
For each utterance stored across the 200,000 files from the previous step, we fetch audio staged in S3 and group them based on speaker. We then sort this audio by its creation timestamp to provide an audio stream per speaker; we extract features for every frame, one speaker at a time. The algorithm also performs a causal mean normalization over all the audio for a speaker. The total size of the features at this stage is around 135 TB. For training deep learning models, features need to be shuffled across the entire file set. We developed a hierarchical shuffling strategy: that both shuffles the global ordering of the files and the features corresponding to the utterances within each of the 200,000 files. We then normalize features using the global statistics: zeroth, first and second order statistics. Statistics for each shuffled file are computed in parallel, and then aggregated across files. Finally, we subsample the features to a lower frame rate (33Hz), while ensuring no information loss by splicing subsampled frames.
\subsubsection{Target generation}
\label{sec:dp_targetgeneration}
We use a trained teacher model to generate training targets for the student model. Target generation is more efficient on GPUs with batching. As the stand alone spark scheduler we used in our data pipeline could not efficiently distribute tasks in multi-GPU machines, we implemented a custom local scheduler for this step to distribute the load on all available GPUs in the cluster.
\subsubsection{Repartition targets and features}
The final step of the data pipeline is to repartition the files to nearly 500,000 partitions (each with about 2 hours of audio) to improve the efficiency of distributed model training.

\subsection{Implementation Considerations}
The pipeline consists of four steps that starting from audio data in S3 executes sequentially to produce inputs for training. Each step is a cluster job that runs on a fixed pool of identical EC2 instances that takes as input, the output from one or more of the previous steps and a cluster configuration. Inputs and outputs are lists of objects in S3. The steps are modular and can be invoked independently given the inputs required for that step. Apart from advantages of modularity and logical separation of concerns, another reason for this design is the large variance in compute resources for individual steps; thus separating the steps based on the computing needs per step allowed us to be more efficient. Cluster configuration for each step includes:

\begin{itemize}
\item EC2 Instance types and the number of instances of that type to be used.
\item S3 resource permissions for read and write operations on S3 objects.
\item Software dependencies for the job, each of which gets deployed on the entire Spark cluster including the master node and the slave nodes.
\item Spark configuration for the job such as the number of executors per node, number of cores to allocate per executor and the Java Virtual Machine(JVM) settings for each executor.
\end{itemize}

\section{The Model pipeline} \label{SSL Model pipeline}
A few design decisions were critical not just for performance, but also for relatively fast experiment turnaround time as well as to be able to build a model on 1 Mhrs efficiently. In this section we report the key ML design choices in our system.

\subsection{Student-Teacher Learning}
We used the student/teacher learning methodology~\cite{jinyu_student_teacher},~\cite{ba2014deep},~\cite{hinton2015distilling},~\cite{million2019icassp} thus simplifying the SSL modeling recipe and eliminating the need for a full ASR decoder. For each feature vector, the teacher and the student networks compute posterior probability distributions over senones. While the teacher parameters are frozen, the student network's parameters are estimated by minimizing the cross-entropy loss between the two posterior distributions. The student network structure is identical to the LSTM AM described in the Section~\ref{Overall System}, but the teacher networks have five bi-directional LSTM layers, each with 768 units (totaling 78 M model parameters) -- this is nearly 3 times the size of the student network. The training of the teacher network on labeled data uses the same recipe as the regular LSTMs.
\subsection{Confidence Modeling}
It has been reported previously that modeling even with unfiltered data can lead to significant WER improvements in the context of SSL (\cite{lamel2002lightly}, \cite{lamel2002unsupervised}). Furthermore, as neural network technology has improved, so have the estimated probabilities become better calibrated~\cite{guo2017calibration}. Our hypothesis is that the teacher's posteriors are well enough calibrated to act as the confidence measure for student network training. However, in a traditional self-training system, the language model also provides additional information during decoding; in our SSL system the LM is not present. We speculate that this is partially mitigated by using bi-directional teacher LSTM models, which observes more context than the student to make a frame level decision.
\subsection{Target Generation}
\label{sec:mp_targetgeneration}
Since the senone output distribution is large (a 3,183 dimensional vector), generating targets using the teacher model on-the-fly slows down training. As we parallelize training across multiple GPUs, to reduce network bandwidth and to minimize storage, we store only the $k$ highest valued logits. During the student network training, full posteriors are reconstructed; the missing logits are filled with large negative values. While the reconstruction procedure is lossy, our experiments showed that probability mass is dominated by a few top posteriors. The hyper parameter $k$ is selected empirically based on the teacher model family, its model structure, and the amount of data it has been trained on.
\subsection{Scheduled Learning}
\label{sec:scheduled_learning}
Although most of the data used for training is unlabeled, we found that using the limited labeled data can be useful in obtaining performance improvements. Our learning algorithm interleaves parameter estimation on unlabeled and labeled data, with slightly higher learning rates on the labeled data. Given the size of unlabeled data, our design was to perform just one epoch through it while visiting the labeled data multiple times. We divided the unlabeled data into a fixed number of \textit{sub-epochs}, with a sub-epoch defined as 55,000 hours. We decayed the learning as we ingested unlabeled data through the sub-epochs, following an exponential learning rate decay. After each sub-epoch through the unlabeled data, we perform CE training on the labeled data, with a rotation through the feature offsets (please refer to Section~\ref{Overall System}). As described in Section~\ref{Overall System} we employ sequence chunked BPTT for training speed; we apply chunked training for the first 15 sub-epochs, and then perform fine-tuning during the last three sub-epochs.
\subsection{Sequence Training for SSL}
Sequence discriminative training of a deep learning AM often yields large WER improvements (commonly, around 10$\%$ relative~\cite{vesely2013sequence}). However, discriminative training is a difficult problem for SSL (\cite{huang2010semi}, \cite{manohar2015semi}), since the discriminative loss function can be sensitive to noisy references during training. Our decision was to perform sMBR training of the student model only on labeled data. Previous work~\cite{kanda2016investigation} indicates that the accuracy gains may be relatively small in such a setup. However, we hypothesized that this result was likely due to a relatively small labeled dataset, and using the full 7,000 hour labeled data in this study could still yield large gains from sequence training.
\subsection{Distributed Training}
Identifying a good approach to performing distributed training of the student model was a key element of our design, exploring the tradeoff between scalability and accuracy. We studied Gradient Threshold Compression (GTC)~\cite{strom2015scalable} and Blockwise Model Update Filtering (BMUF)~\cite{bmuf-2016}, of which we discuss more in Section~\ref{Distributed Training}.

\section{Distributed Training} \label{Distributed Training}

For large-scale neural network training, distributing the workload across many GPUs is required to produce a trained model in a reasonable time. Here we will discuss data-parallelism, where different worker nodes process different input data, but share the model that is trained. The widely used stochastic gradient descent (SGD) opimization technique (e.g. \cite{hintondeep:2012}, \cite{convasr}) has a serial aspect to it that makes it challenging to scale SGD to large number of workers. Let's assume that the standard technique of using a ``mini-batch" is used. A mini-batch is a small batch of data-points that are processed efficiently on a single GPU. Basic data-parallelism then involves computing mini-batches on all workers, aggregating their gradients and updating the shared model. However, by increasing the number of compute nodes, the effective (aggregated) mini-batch size is increased linearly, which has shown to produce lower accuracy on the validation and test datasets, and generally reduces model convergence rate \cite{imagenet1hr}, \cite{talben}. Several techniques can be employed, such as adjusting learning rate \cite{imagenet32k}, \cite{imagenet1hr}, using a warm-up phase (e.g., \cite{strom2015scalable}), etc. which can increase an upper bound on the workable effective mini-batch size, but fundamentally there still remains an obstacle of scaling to large GPU clusters.

A specific scaling challenge is communicating gradients between workers which requires high bandwidth and as the size of model or the number of workers are increased, this can lead to a severely communication-bounded algorithm. This also depends on whether workers are on the same host or different, since the cost of communication between different hosts is typically much higher. In case of larger number of workers, the cost of communication can thus dominate the total training time. In particular for cases where the ratio of compute-time to communication is low, having high bandwidth interconnect is then necessary to reduce overall time spent in communication. Empirically we find that with high-end compute nodes such as AWS p3.16xlarge which contains 8 Nvidia V100 GPUs and provides data transfers upto 300 Gbps across peer GPUs on the same host, data parallel SGD can scale almost linearly within a single host. Algorithms such as ring-allreduce \cite{Patarasuk2009BandwidthOA} and hierarchical ring-allreduce \cite{hallreduce} are used which aim to utilize available bandwidth optimally among compute devices. However due to the limited bandwidth of 25 Gbps across hosts on these instances, the scalability of the training beyond single host is severely affected.

Several techniques have been developed to reduce the bandwidth required in gradient communication. The early works of \cite{1bit-sgd} and \cite{strom2015scalable}, that introduced the quantization and compression by gradient thresholding, have later been refined and used in many contexts, e.g. \cite{qsgd}, \cite{deepcompression}, \cite{adacomp2017}. These techniques reduce the amount of data which needs to be communicated between workers to reduce overall communication time. This increases the limit on the number of GPU worker nodes that can be used in parallel without rendering the algorithm communication-bound.

Another well-known general technique to scale distributed training is based on the concept of Model Averaging (MA). In this technique each worker updates a local model based on many mini-batches from the dataset without communicating with other workers. The model is only synchronized across workers after some interval of time or specified number of mini-batches. At that time, all model weights are averaged across workers and synchronized. Because of the infrequent synchronization, this approach can scale training data throughput almost linearly. However in its basic form, it suffers from reduced model training convergence because non-linear divergence between local models that is not well-matched with averaging. A variant, Blockwise model update filtering (BMUF) \cite{bmuf-2016}, mitigates this issue significantly.

In this paper we are comparing and contrasting one method from each of the above mentioned data parallel approches, namely gradient averaging and model averaging. We selected synchronous SGD with gradient threshold compression\cite{strom2015scalable} (GTC) and BMUF \cite{bmuf-2016} which both try to address the issue of scaling data parallel training.

\subsection{Gradient Threshold Compression}
In GTC, instead of sending the entire gradient tensor for each trainable weight, only gradient elements whose absolute magnitude is greater than a constant, here referred as gradient-threshold ($\tau$) are sent to other workers. This results in a very sparse gradient update -- typically reducing the gradient size by several orders of magnitute. Each worker communicates the sparse update to all other workers and conversely receives all sparse updates from other workers. The received sparse gradient updates are aggregated and weights are updated based on the aggregate. The residual gradients which are not sent to other workers are aggregated locally for later iterations. In a naive implementation, a sparse update can be represented by two numbers, an integer element index and floating point number. However this can be compressed further by quantizing gradient and packing quantized gradient and integer index into single 32-bit integer field. In this work, we use 1-bit quantization \cite{1bit-sgd}. Thus, each worker simply sends gradient deltas of $\pm\tau$. This simple coding scheme further compresses the update by 2x. The pseudo-code for this algorithm and a more comprehensive discussion can be found in \cite{strom2015scalable}. The technique can be applied to synchronous as well as asynchronous variants of SGD, however we select the synchronous variant to ensure reproducibility.

\subsection{Blockwise Model Update Filtering}
The BMUF algorithm \cite{bmuf-2016} is a variant of model averaging, augmented by considering the model from previous step. First the initial global model ($W_g$) is broadcasted to all workers. The algorithm then iterates two main steps. In the first step, each worker updates its local model ($W$) in parallel with its portion of data for a specified number of mini-batches, here referred as block-size. This step is called intra-block parallel optimization and requires no synchronisation between workers. In our implementation, each worker simply updates its local model using mini-batch SGD independently. In the second step, which is referred to as the BMUF step, the global model is updated using the following procedure.

\begin{equation} \label{eq:MA}
\overline{W}(t) = \frac{1}{N} \sum_{i = 1}^{N} W(t)^i
\end{equation}
\begin{equation} \label{eq:Gt}
G(t) = \overline{W}(t) - W_g(t-1)
\end{equation}
\begin{equation} \label{eq:Delta}
\Delta(t) = \eta_t\Delta(t-1) + \zeta_tG(t)
\end{equation}
\begin{equation} \label{eq:NBM}
W_g(t) = W_g(t-1) + \Delta(t) + \eta_{t+1}\Delta(t)
\end{equation}

where hyper-parameters $\eta$ and $\zeta$ are called block-momentum and block-learning-rate respectively. We used following formula
\begin{equation} \label{eq:BMUFC}
\frac{\zeta}{N(1-\eta)} = C
\end{equation}
to set $\eta$ and $\zeta$ hyper-parameters, where $C \geq 1$ is constant and $N$ is number of workers. We use Nesterov block momentum (NBM) scheme proposed in \cite{bmuf-2016}.

The evaluation of these two training methods for frame level accuracy and speedup is described in section \ref{sec:dist_train_results}.

\subsection{Accuracy and Scalability Trade-offs}
Both above mentioned algorithms provides flexibility to scale to large number of workers through hyper-parameters, however it may come at the cost of reduced accuracies when number of workers are large. The GTC algorithm can be scaled by controlling gradient-threshold parameter which directly affects sparsity of gradient values. This results in lower update size and reduces overall communication time. The trade-off of different gradient-threshold values on accuracy and speed is described in \cite{strom2015scalable} for asynchronous variant of the algorithm. In our studies we found gradient-threshold of 8 achieved best trade-off between accuracy and scalability. For the BMUF algorithm, the block-size hyper-parameter can be used for controlling how often global model is updated. Setting block-size to large number can enable almost linear scaling, however this results in considerable drop in accuracy for large number of workers. This is further analyzed in \cite{bmuf-2016}. In our studies, we found block-size of 100 achieved best trade-off between accuracy and speed.

\section{Experimental Setup} \label{Experimental Setup}
We discussed system level details in Section~\ref{sec:sys_description}. In this section we provide details on our experimental setup, including the training and test data sets. We also discuss various models, and briefly describe the decoding setup.

\subsection{Training Datasets}
For our experiments we used three far-field training datasets drawn from the production data of the Alexa family of devices from the US English locale: (a) a 1,000 hour fully labeled dataset for distributed training experiments, (b) a 7,000 hour fully labeled dataset used for training the teacher model, and (c) a 1 Million hour unlabeled dataset for SSL model build. The 1,000 hour dataset is a subset of the 7,000 hour dataset.

\subsection{Test Datasets}
We used two test sets in this work. The first is a validation test set (referred to as VAL), which consisted of about 1 hour of data to evaluate the distributed trainers. The accuracy on this test set is evaluated using frame classification accuracy, but more importantly, we use it to measure training speed. The second test set (TST) consists of audio data collected in a real room with about 5,000 utterances roughly equally spread among five device placements. The first device placement (DP1) in the center of the room led to the lowest error rate, while other conditions (DP2 to DP5) were more challenging. On this test set we use word error rate reduction (WERR) to evaluate the model performance. We have also reported on other test sets in ~\cite{million2019icassp}.

\subsection{Models}
In this section we summarize the models relevant to the experimental setup. All the acoustic models (recognition and alignment models) employed in this paper use the hybrid HMM-LSTM approach.
\subsubsection{Acoustic front-end}
\label{sec:frontend}
The sampling rate of the speech signal in all the datasets used in this work is 16 kHz. The features for the deep learning models come from an acoustic front-end that outputs 64 dimensional log filter bank features at a frame rate of 33 Hz; section Section~\ref{sec:sys_description} describes it in greater detail. The phonetic decision tree, however, was built using 40 dimensional features from a different front-end: application of LDA followed by MLLT transforms on 39 dimensional PLP, including delta and delta-delta.
\subsubsection{Frame level hard targets}
The triphone HMM states were clustered down to 3,183 senones using a phonetic decision tree built on the 7,000 hour dataset. Alignments from the alignment model were mapped, and several rounds of realignment followed by parameter reestimation were performed. Using these alignments, an LSTM trained using the CE criterion, discussed in Section~\ref{sec:sys_description} is used to generate frame-level targets for training the supervised models in this work. All the alignment models used the GTC trainer in conjunction with 16 V100 GPU cards.
\subsubsection{Sequence training}
Sequence discriminative training in this paper used the sMBR~\cite{kingsbury2009lattice} criterion with lattice based methods. sMBR training, for all models including the teacher and the student models, was performed on the fully labeled 7,000 hour dataset. It used GTC trainer with 16 V100 GPU cards. The lattices themselves were shallow, with an average density of around 10, and were stored as compressed files. The space required for storing the lattices for a system was around 6 GB.
\subsubsection{Teacher model}
The teacher is a bidirectional LSTM (BLSTM) model built on the 7,000 hour fully labeled dataset using the features and frame-level targets discussed in this section. This model has 5 bidirectional LSTM layers, each with direction and layer having 768 units. The model parameters were first estimated by minimizing the frame level cross-entropy criterion. The training strategy discussed in Section~\ref{sec:sys_description} was followed: 10 epochs of chunked training, followed by 2 epochs of fine-tuning. Finally 3 epochs of sequence training was performed.
\subsubsection{Features and targets for the student model}
Features for the 1 Million hour dataset were generated using the system described in Section~\ref{sec:dp_steps}. As in Section~\ref{sec:frontend}, this results in 64 dimensional log filter bank features at a frame rate of 33 Hz. Using the trained BLSTM teacher model, frame level soft targets are generated using these features and stored as compressed $k$-best logits (with $k$ being 20 for the experiments) using the techniques discussed in Sections~\ref{sec:dp_targetgeneration} and~\ref{sec:mp_targetgeneration}.
\subsubsection{Student model}
The student network is identical to the LSTM architecture described in Section~\ref{sec:sys_description}. The student model is trained on the features and targets discussed in the previous subsection with scheduled learning, as discussed in Section~\ref{sec:scheduled_learning}. We used the BMUF trainer with 8 p3.16xlarge hosts. Lastly 3 epochs of sMBR training restricted to 7,000 hours with GTC trainer was performed.
\subsubsection{Baseline fully supervised model}
\label{sec:fully_supervised}
The baseline fully supervised system is an LSTM, identical to the network discussed in Section~\ref{sec:sys_description}. This network was trained as discussed in Section~\ref{sec:sys_description}, on the 7,000 hour dataset, using the same set of features and targets used for training the teacher model. Lastly, 3 epochs of sequence training was performed on the same labeled dataset.

\subsection{Decoding Setup}
All decoding on the TST test set use a 4-gram statistical language model (LM). The acoustic model scale factor was tuned on this test set. We compare the SSL model against a strong, fully supervised baseline system~\cite{million2019icassp}.

\section{System Evaluation} \label{Results}
We evaluate the system along several dimensions. A key metric is the accuracy of the trained models. For this dataset, accuracy is reported as relative word error rate reduction (WERR) (cf. ~\cite{million2019icassp}, ~\cite{garimella2015robust}, ~\cite{king2017robust}, ~\cite{parthasarathi2015fmllr}, ~\cite{minhua2019icassp}). Here we show a subset of accuracy results with a more complete picture available in ~\cite{million2019icassp}. Other important metrics are the processing time and cost of infrastructure.

\subsection{Distributed Training}
\label{sec:dist_train_results}

We trained models using GTC and BMUF on the 1,000 hour labeled data\footnote{We used AWS EC2 p2.16xlarge instances for experiments in Table~\ref{tab:speed}. where each instance consists of 16 NVIDIA K80 GPUs.}, which were then evaluated on the VAL test set.

The training speedup and relative frame-level classification accuracy improvements of the two trainers are tabulated in Table~\ref{tab:speed}. Both metrics are relative to a 1-GPU SGD trainer which does not perform any gradient thresholding or quantization. It can be seen that in terms of accuracy, both trainers are within 1$\%$ relative compared to the 1-GPU SGD baseline. BMUF trainer shows higher degradation of about 3$\%$ relative when run on 128 GPUs. Synchronous GTC training scales well up to 32 GPU cards, but its throughput tapers off at higher scale. This is due to increased cost of communication needed to synchronize gradients per mini-batch as number of workers are increased. On the other hand, BMUF scales almost linearly with number of workers, at least in terms of throughput, since model synchronization is much more infrequent. However, it comes at a cost of training convergence rate and reduced accuracy at higher number of workers. The Nesterov-like momentum updates at block level recover some of these losses, but empirically we still see some degradation.

%\begin{small}
\begin {table}
\caption {\textit{Relative frame-level classification accuracy improvements (in $\%$) and training speedup (as a factor) of GTC and BMUF-NBM compared to 1-GPU SGD trainer. This table illustrates the trade-offs for the two trainers, as a function of the amount of compute (number of GPUs).}} \label{tab:speed}
\begin{tabular}{ | P{1.8cm} | P{1.6cm} | P{2cm} | P{1.4cm} | }
\hline
\rule{0pt}{2ex} Training Method & Number of Workers & Relative Frame-level Accuracy Improvement (\%) & Training Speedup \\[2pt] \hline
\multirow{5}{*}{\parbox{2cm} {GTC}} & 8 & 0.41 & 7.01 \\ \cline{2-4}
 & 16 & 0.41 & 12.53 \\ \cline{2-4}
 & 32 & 0.54 & 21.75 \\ \cline{2-4}
 & 64 & 0.27 & 16.76 \\ \cline{2-4}
\hline \hline
\multirow{7}{*}{\parbox{2cm}{BMUF-NBM\\ block-size: 100}} & 8 & -0.27 & 7.18 \\ \cline{2-4}
 & 16 & -0.06 & 13.34 \\ \cline{2-4}
 & 32 & -0.10 & 25.46 \\ \cline{2-4}
 & 64 & -0.13 & 50.91 \\ \cline{2-4}
 & 128 & -2.46 & 97.59 \\ \cline{2-4}
\hline
\end{tabular}
\end {table}
%\end{small}

\subsection{End-to-End Processing Time}
The end-to-end time from data to a fully trained model yields an assessment of our system design. A breakdown of the processing times also gives the constraints that limit our ability to scale the model training to even larger data regimes.
\subsubsection{Proposed system design}
\begin{figure}[htb]
 \centering
\includegraphics[width=4.5in]{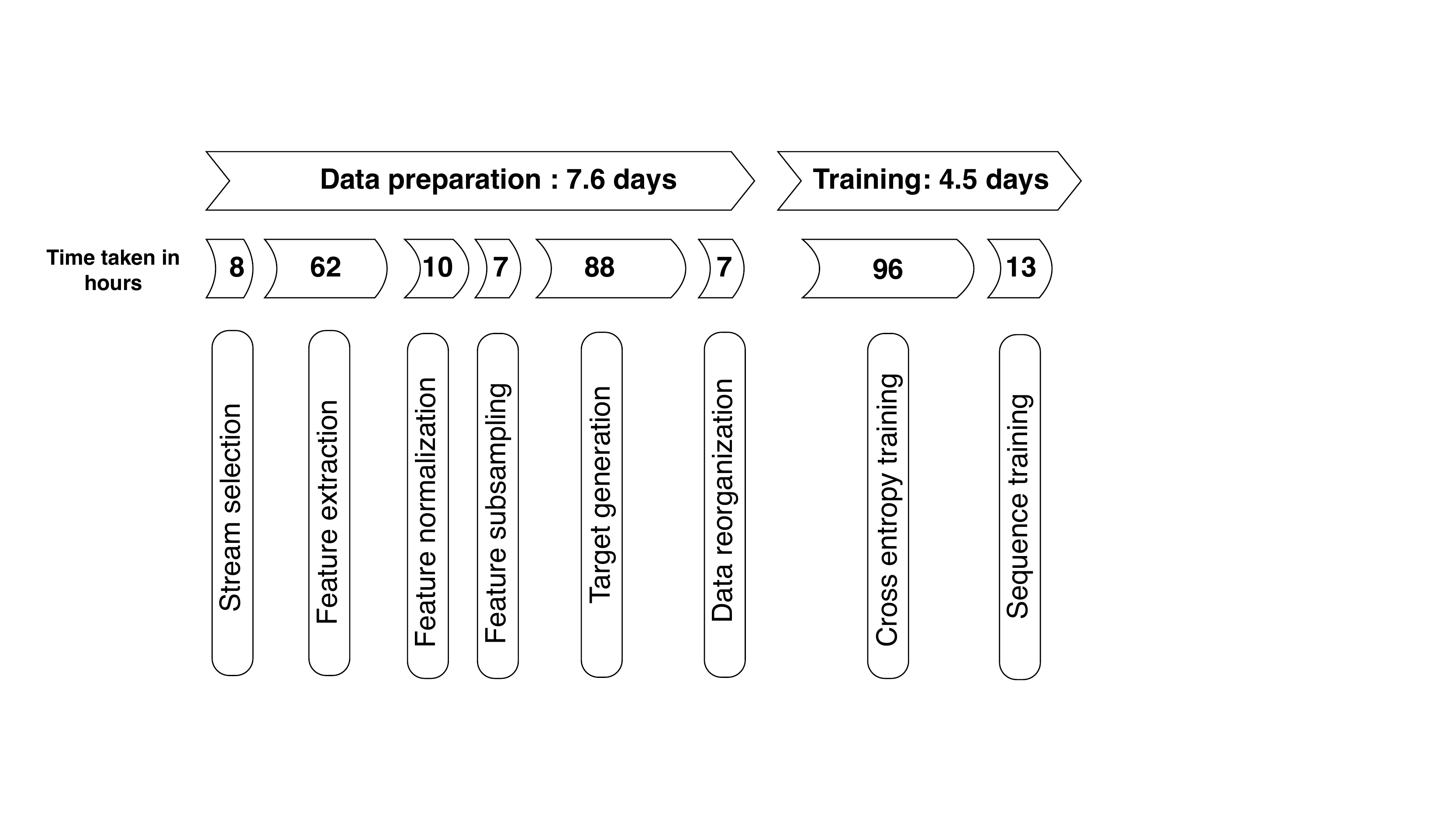}
 \vspace{-1cm}
 \caption{\textit{This figure presents a breakdown of the end-to-end processing times to train a full SSL student subsystem. The data pipeline scales linearly with more compute and can have even quicker turnarounds; using more compute, the model training can scale further in terms of training times, but it comes at the cost of accuracy.}}
 \label{fig:evaluation3}
\end{figure}

With the design decisions taken in this work we obtain an end-to-end turn around time of the student pipeline in 12 days. Figure~\ref{fig:evaluation3} breaks down the processing time in different parts of the student training pipeline. As a side-note, the initial training of the teacher model and the storing of utterances corresponding to the 1 Million hours of speech in S3 takes an additional 4 days (the training of the teacher model itself takes 2 days\footnote{It might be surprising that the training of the teacher model on 7,000 hours takes 2 days, while the training of the student model takes ``only'' 4.5 days; but the teacher is a BLSTM, using GTC trainer employing 16 GPUs. Also we perform 12 epochs of training on 7,000 for the teacher. The student model is an LSTM (trains 2x faster), using BMUF trainer employing 64 GPUs (about 4x faster), doing 1 epoch through the data.}).

The SSL data pipeline for generating features and targets for the 1 Million hours of speech took 7.6 days. This pipeline is relatively straight-forward to parallelize. For most steps, since each data partition is independent, adding more hardware can parallelize the step further. We perform distributed computation on Spark, and all our data is staged in S3. Since both systems are known to scale well, the data pipeline can scale nearly linearly by increasing the cluster size. Further algorithmic improvements are possible with more caching, pre-computations and aggregations, though at the cost of more storage. Model training contributes to a smaller part of the total time (4.5 days). In this project, we increased the number of GPUs to 64, for a very significant speed-up. Adding further compute by using more GPUs can speed up training linearly, but has diminishing returns in terms of accuracy (as presented in Table~\ref{tab:speed}).

\subsubsection{Comparison to fully supervised system}
We close the discussion on processing times by making a note on the fully supervised AM described in Section~\ref{sec:sys_description}, and the model was built as presented in Section~\ref{sec:fully_supervised}. This involved 2 steps:
\begin{enumerate}
\item \textit{Feature extraction and target generation}: This was the bottleneck: this system was impractical for feature extraction and target generation at the scale of 1 Million hours. For the 7,000 hour data, the feature and target generation took nearly 2 weeks.
\item \textit{Model building including the CE and sMBR training stages}: Model building was done with GTC trainer using 16 GPU cards for both the CE and sMBR stages. This took about 21 hours.
\end{enumerate}

\subsection{System Accuracy}
The final results including sequence discriminative training, with sMBR loss function, are reported in Table~\ref{tab:tab_final_1} on the TST test set. We compare against the baseline fully supervised model trained on the 7,000 hour data (please refer to Section~\ref{sec:fully_supervised} for more details on this model). The results are reported as relative WER improvements.
\begin{table}[h]
\centering
\caption{\textit{On TST test set (in DP1 to DP5), relative WER reduction ($\%$) of the final 1 Million hour model against a baseline LSTM AM that is sMBR trained on the fully labeled 7,000 hour training data.}}
\renewcommand{\arraystretch}{1.3}
\begin{tabular}{|c|c|c|c|c|c|}
  \hline
Test Conditions & WERR ($\%$) \\
 \hline
DP1 & $9.8$  \\
\hline
DP2 & $22.2$ \\
\hline
DP3 & $21.8$  \\
\hline
DP4 & $16.5$   \\
\hline
DP5 & $18.9$   \\
\hline
\end{tabular}
\vspace{-4mm}
\label{tab:tab_final_1}
\end{table}

Except for device position one (DP1), relative WER reductions are all greater than 10$\%$, and indicating that the improvement is greater for harder conditions. We take this as validation that large scale SSL can not only significantly improve accuracy overall, but also yield significant improvement in more challenging conditions.

\section{Conclusion} \label{Conclusion}
In this paper we presented an in depth discussion of the design of an efficient end-to-end SSL system starting from 1 Million hours of raw audio and its metadata. Following the student/teacher paradigm for SSL, we focused on the student subsystem, factoring it into two main pipelines: data preparation and model training. To address the challenges of scalability and robustness, our discussion on data pipeline laid out the key bottlenecks and proposed corresponding design principles. These principles were then used in decomposing the pipeline into smaller steps to efficiently address the challenges.

The model pipeline, including the distributed trainer, addressed the twin challenges in ML design for this problem: accuracy improvement and scalability. Scaling posterior generation with k best selection, using scheduled learning to leverage transcribed data, restricting sequence training to transcribed data were among the methods we presented. Our system evaluations showed that even without extensive hyper-parameter tuning, we can obtain relative WER improvements in the 10 to 20$\%$ range, with much higher gains in more difficult conditions. The end-to-end processing time of this SSL system was 12 days, and several components in this system can trivially scale linearly with more compute resources for further speed-up.

\section*{Acknowledgment}
We would like to thank Xing Fan for providing the setup for baseline models; Harish Mallidi for help with the decoding infrastructure; Oleg Rybakov and Tianjun Ye for help with early debugging with our deep learning toolkit.

%\bibliographystyle{IEEEtran}
%\bibliography{combined}

\nocite{*} % to test all bib entrys

% === Pasted from ssl_jrnl.bbl ===


% ================================


%\vfill


\end{document}